# Leech Lattice Extension of the Non-linear Schrodinger EquationTheory of Einstein Spaces

George Chapline
Lawrence Livermore National Laboratory

Although the nonlinear Schrodinger equation description of Einstein spaces has provided insights into how quantum mechanics might modify the classical general relativistic description of space-time, an exact quantum description of space-times with matter has remained elusive. In this note we outline how the nonlinear Schrodinger equation theory of Einstein spaces might be generalized to include matter by transplanting the theory to the 25+1 dimensional Lorentzian Leech lattice. Remarkably when a hexagonal section of the Leech lattice is set aside as the stage for the nonlinear Schrodinger equation, the discrete automorphism group of the complex Leech lattice with one complex direction fixed can be lifted to continuous Lie group symmetries. In this setting the wave function becomes an 11x11 complex matrix which represents matter degrees of freedom consisting of a 2-form abelian gauge field and vector nonabelian SU(3)x$E_6$ gauge fields together with their supersymmetric partners. The lagrangian field equations for this matrix wave function appear to be a promising starting point for resolving the unphysical features inherent in the general relativistic descriptions of gravitational collapse and the big bang.

## Introduction

Among the known exact solutions of the Einstein's general relativity equations self-dual Einstein spaces stand out as especially interesting [1]. Our interest is these spaces was stimulated by the discovey that these solutions can also be expressed as solutions of a matrix nonlinear Schrodinger equation in 2+1 dimensions with an SU(N) Chern-Simons gauge potential [2]. This in turn led to a "superfluid" representation for the vacuum state of a space-time with no matter [3,4]. Independently of this development there are also quite general reasons [5,6] for supposing that in a quantum theory of gravity space-time can be descibed as a superfluid of "gravitons". It is encouring in this connection that, in contrast with the prediction of general relativity, there are both theoretical [7-9] and phenomenological [10] hints that in a quantum theory of gravity the surface of compact objects is not an event horizon, but instead represents a phase transtion in the vacuum state of space-time [11]. If it turns out that the general relativisitic description of compact objects is indeed flawed, the fault may well lie the likelihood that a consistent treatment of space-times with matter will require a quantum theory of space-time. Unfortunately, a quantum theory of space-times with matter has remained elusive up to the present time.

In this paper we wish to draw attention to the possibility that a matrix non-linear Schrodinger equation living on a Lorentzian extension of the 24-dimensional Leech lattice may provide the needed framework for a quantum theory of gravity and elementary particles in 3+1 dimensions. Our basic idea is to replace the complex scalar wave function and gauge potential used in ref's [2,3] to encode the Kahler potential of a self-dual Einstein space by complex matrices representing discrete symmetries of the complex Leech lattice. The Leech lattice was originally constructed as a result of its connection with perfect error correcting codes [12,13]. From the point of view of constructing a quantum theory of space-time containing matter perhaps the most intriguing property of the Leech lattice is that there is a correspondence between the shapes of the minimal Leech lattice vectors (the set of such vectors centered on a point of the lattice will hereafter be referred to as the "Leech polytrope") and the types of massless fields that occur in the 10-dimensional unification of supergravity and super Yang-Mills theories [14]. For example, the number of vectors of the Leech polytrope with shape $(4^2, 0^{22})$ is equal to 4x the number of components of a 2-form field $B_{\mu\nu}$ in 24-dimensions, while the vectors with shape $(-3, 1^{23})$ represent 2x the number of components of a gravitino field in 24 dimensions. An interesting feature of the Leech lattice is that if the Leech lattice generators with shape $(8, 0^{23})$ – which are not vertices of the Leech polytope - are added to the vertices of the Leech polytrope, then the numbers of bosonic and femionic degrees of freedom are exactly equal. Remarkably the full automorphism group of the Leech lattice, the Conway group 0• [15], includes supersymmetry-like transformations between Leech polytrope vectors representing bosonic and fermionic degrees of freedom. Thus our Leech lattice set-up parallels the classical unification of supergravity and super-Yang-Mills interactions in 10-dimensions, with the caveat that the gravitational degrees of freedom are only active in the 3+1 dimensions of ordinary space-time.

In order to breathe life into the degrees of freedom associated with the Leech lattice, the wave function and gauge potential will be allowed to vary with position within a distinguished 2-dimensional hexagonal section of the Leech lattice. When time is added to the nonlinear matrix equation the wave function will depend not only on position within the distinguished 2-dimensional section of the Leech lattice, but also time. A remarkable feature of our set-up is that if the discrete symmetries of of the Leech polytrope are restricted to the sublattice consisting of vectors not lying in the distinguished 2-dimensional section, then some of these discrete symmetries can be extended to Lie group symmetries and one can introduce local gauge invariance for the Schrodinger equation in way compatible with the previous theory of Einstein spaces. Furthermore if a 2-dimensional self-duality condition is imposed on the wavefunction and a 2+1 dimensional Chern-Simons constraint is imposed on the gauge potential, then the time independent matrix nonlinear Schrodinger equation can be solved analyticaly, yielding holomorphic and anti-holomorphic solutions that carry internal degrees of freedom in addition to the usual gravitational degrees of freedom in 2+1 dimensions. If these solutions are paired to provide a "superfluid" wave function for space-time, the low lying excitations of the ground state will resemble familiar massless gravitational, anti-symmetric tensor, and supersymmetric Yang-Mills degrees of freedom.

As will be discussed in section 3, the degrees for freedom in our model not directly related to the metric for 3+1 dimensional space-time are related by an especially interesting discrete symmetry, $M_{11}$, acting on the vectors of the complex Leech polytrope. The $M_{11}$ group was discovered in the 19[th] century by the mathematical physicist Emil Mathieu, and was the first in a series of discoveries of sporadic finite simple groups that culminated in 1981 with Griess' construction of the Monster sporadic group. [13,15]. Perhaps the most remarkable aspect of our model for space-times with matter is that provides a link between the unification of gravity and elementary particle physics and the mathematical structure of the sporadic finite simple groups including the Moster. That this should be so is of course another illustration of Wigner's "unreasonable usefulness of mathematics in physics".

Another notable feature of our model for space-time is that the vacuum energy density will in general not be zero. Actually the numbers of bosonic and fermionic degrees of freedom for the 2-form and Yang-Mills matter fields in our theory exactly match. This means that at the level of 3+1 dimensions the contribution of zero point fluctuations of the matter fields to the ground state energy will vanish. However, the contibutions of zero point fluctuations of the bosonic and fermionic gravitational degrees of freedom to the ground state energy will not exactly cancel each other. In addition, the direct interaction of self-dual and anti-self-dual solitons in our quantum vacuum state will produce a nonzero energy density.

In Section 2 we review the gauged nonlinear Schrodinger equation theory that provides a quantum model for classical Einstein spaces. In Section 3 we exhibit the gauged nonlinear Schrodinger equation that we believe provides a framework for understanding the quantum nature of space–times wih matter. We also exhibit a matrix generalization of the classical "Heavenly" equation for the Kahler potential for a self-dual or anti-self dual Einstein space, whose soliton-like solutions provide the building blocks for a theory of space-time with 3-form vacuum fields.

## 2. Non-linear Schrodinger equation for Einstein spaces

The coherent state wave function for a 2-dimensional quantum fluid of anyons interacting via Chern-Simons gauge potentials satisfies the non-linear Schrodinger equation [2-4]:

$$i\hbar \frac{\partial \psi}{\partial t} = -\frac{1}{2m} D^2 \psi + eA_0 \psi - g|\psi|^2 \psi \quad , \tag{1}$$

where $D_\alpha = \partial_\alpha - i(e/\hbar c)A_\alpha$ and $m$ is a mass parameter. The gauge fields $A_0$ and $A_\alpha$ do not satisfy Maxwell's equations, but instead are determined self-consistently from the equations for Chern-Simons electrodynamics in 2+1 dimensions. In the presence of a uniform 2-dimensional electric field $E$ the current has the same form as the Hall current for a magnetic field perpendicular to the plane.

$$j_{\alpha\beta} = \sigma_H \varepsilon_{\alpha\beta\gamma} E_\gamma, \qquad (2)$$

where $\sigma_H$ is the "Hall conductivity". Neglecting spatial variations in the electric field, the usual Guass' law will be replaced by the Chern-Simons equation

$$B = -\frac{\rho}{\kappa}, \qquad (3)$$

where $B$ is the strength of an effective magnetic field whose direction is perpendicular to the layer, $\rho$ is the charge per unit area, and $1/|\kappa|$ is an inverse length with $\sigma_H = \kappa$. It is interesting that these equations also appear in a tight binding model for the motion of charged particles on a hexagonal lattice [16]. It was shown some time ago [17] that the time independent version of Eq. (1) in conjunction with Eq's (1-2) can be solved analytically if one assumes that

$$g = \pm e^2 \hbar / mc\kappa. \qquad (4)$$

The ground state solution of Eq. 1 contains solitons with vortex-like currents and two units of quantized magnetic flux attached to every carrier. The two signs for $g$ correspond to solutions where the vorticity of all the solitons is either up or down. These equations

In 1991 the author and Kengo Yamagishi [2] introduced the idea of extending the nonlinear Schrodinger equation in 2+1 dimensions to 3+1 dimensions by replacing the 2-dimensional complex plane with a stack of N such planes and the scalar wave function with an SU(N) matrix:

$$i\hbar \frac{\partial \Phi}{\partial t} = -\frac{1}{2m} D^2 \Phi + e[A_0, \Phi] - g[[\Phi^*, \Phi], \Phi] ,  \quad (5)$$

where the wave function $\Phi$ and potentials $A_0$ and $A_i$ are now SU(N) matrices, and $D \equiv \nabla - i(e/\hbar c)[A,$. Promoting the scalar wave function ψ(z) to an SU(N) matrix wave function Φ(z) is a way to take into account into account inter-layer interactions and tunneling. The matrix magnetic field $B_{\text{eff}} = \partial_x A_y - \partial_y A_x + [A_x, A_y]$ seen by charged particles in 2-dimensions is now a diagonal matrix

$$B_{\text{eff}} = -\frac{e}{\kappa}[\Phi^*, \Phi] . \quad (6)$$

The in-plane electric field $E_\alpha$ will also be a diagonal matrix satisfying the Hall effect equation::

$$E_\alpha = -\frac{1}{\kappa} \varepsilon_{\alpha\beta} j_\beta , \quad (7)$$

where $j_\alpha = (\hbar/2mi)([\Phi^*, D_\alpha \Phi] - [D_\alpha \Phi^*, \Phi])$ is the in-plane current. Time independent analytic solutions to eq. (5-7) can be found for any value of N if Eq.(4) is satisfied. These analytic solutions satisfy the 2-dimensional self-duality condition $D_\alpha \Phi = \pm i\varepsilon_{\alpha\beta} D_\beta \Phi$ and represent zero energy ground states for a stack of N planes (or N hexagonal lattices).

In the limit N → ∞ the analytic solutions of Eq. 5 take a particularly simple form such that the effective magnetic field seen by the jth soliton has the simple form:

$$B_j = \pm \frac{\hbar c}{e} \sum_k \nabla_k |X_j - X_k| , \quad (8)$$

where $X \equiv (z, u)$ is now a 3-dimensional coordinate encoding both the position $z = x+iy$ of a soliton within a layer and the height $u$ of the layer. In this solution the vortex-like solitons present in the solution for a single layer have become monopole-like objects, which were christened "chirons" in Ref. 2. The ground state corresponding to (8) has zero energy and the wave function has the form [3,4]

$$\Psi = f(w) \prod_{k>j}^{\infty} \left[ \frac{R_{jk} + U_{jk}}{R_{jk} - U_{jk}} \right]^{1/2}, \qquad (9)$$

where $R_{jk}^2 = U_{jk}^2 + 4(z_j - z_k)(\bar{z}_j - \bar{z}_k)$, $U_{jk} = u_j - u_k$, and $f$ is an entire function of the $\{\bar{z}_i\}$ in the self-dual case and $\{z_i\}$ in the anti-self-dual case. Writing the product on the r.h.s. of Eq. (9) as exp(S) defines an effective action for a gas of chirons:

$$S = \frac{1}{2} \sum_j \ln \frac{R_j + u - u_j}{R_j - u + u_j}, \qquad (10)$$

where $R_j^2 = (u - u_j)^2 + 4(z - z_j)(\bar{z} - \bar{z}_j)$. The wave function (9) resembles in some respects Laughlin's wave function for the fractional quantum Hall effect; for example moving the $z$ coordinate of a chiron around the position of another chiron in a different layer changes S by $i\pi$ [2,3]. However, in contrast with the fractional quantum Hall effect, there are two distinct degenerate ground states corresponding to the self-dual and anti-self-dual solutions for eq. (5), reminiscent of the Kramers pairs in systems with time reversal symmetry. It was the motivation for the suggestion [3] that these two solutions can be combined to yield a model for empty space-time.

Actually the effective action (10) for chirons suggests a connection with the Kosterlitz-Thouless condensation of vortex and anti-vortex pairs in the 2-dimensional XY model [18]. It is an elementary identity that the right hand side of (9) can be rewritten in the form

$$S = \sum_i \pm \tanh^{-1}\left(\frac{u - u_i}{R_i}\right), \quad (12)$$

which is similar in form to a configuration of 2-dimensional XY vortices. In the XY model the phase variations in a 2-dimensional condensate can be described by a partition function of the form

$$Z = \int_0^{2\pi} D\Theta \exp[-\frac{K}{2}\int d^2\xi \frac{\partial \Theta}{\partial \xi_i} \frac{\partial \Theta}{\partial \xi_i}], \quad (13)$$

where $\Theta$ is a periodic coordinate whose period is $2\pi$ and K is a constant. It can be shown that a discrete version of this theory interpolates between the low and high temperature phases of the XY model. Indeed evaluating the exponential in (13) for a configuration of vortices yields the partition function for a 2-D Coulomb gas. On the other hand substituting the chiron effective action (12) into the exponential in (13) yields:

$$\exp-\pi K\left[\sum_{i \neq j} m_i m_j \ln \frac{R_{ij}}{|z_i - z_j|}\right] \quad (14)$$

Expression (14) illustrates why Einstein spaces such as flat Minkowski space-time can also be viewed as a condensation of self-dual and anti–self-dual chirons. Although the pairing of self-dual and anti-self-dual solutions of the Einstein equations must classicaly be defined using two seprate 2-dimensional spaces (the "ambi-twister" construction of Einstein spaces), as first pointed out in Ref. 3 it is also possible to regard (14) as defining a coherent quantum state consisting of pairs of self-dual and anti-self dual solitons. Although this construction specifically applies to space-times with no matter, we will argue in the fillowing that a very similar construction may be used to construct space-times containing certain kinds of vacuum fields.

## 3. Extension to include matter

In this section we will outline how the "superfluid" description of Einstein spaces introduced in Section 2 might be extended to space-times with matter consisting of certain kinds of massless elementary particles or vacuum fields. Since the solutions of the 2-dimensional nonlinear Schrodinger equation that were used to construct self-dual Einstein spaces were either holomorphic or anti-holomorphic functions, we will assume that it is actually the complex Leech lattice, which is a 12-dimensional lattice whose coordinates are Eisenstein integers, rather than the 24-dimensional Leech lattice with real coordinates that is the natural setting for our theory.

Our construction begins by replacing the complex plane used in the construction of Einstein spaces with a distinguished hexagonal lattice section of the Leech lattice. This move is partly motivated by the remarkable observation [16] that the motion of charged particles on a hexagonal lattice naturally gives rise to the Chern-Simons electrodynamics Eq's (2-3). The particular hexagonal section we will use corresponds to the Eisenstein direction $((3\pm\sqrt{3}i)/2)(\sqrt{3}i, 0^{11})$. The automorphism group of the Leech lattice which fixes one point of the lattice, the Conway group $0\bullet$, will be broken by our setting aside this hexagonal section of the Leech lattice. The subgroup of $0\bullet$ which maps the Leech polytrope into itself is $2^{12}M_{24}$ where $2^{12}$ is a group of reflections and other obvious involutions of the Leech lattice and $M_{24}$ is one of the exceptional permutation groups discovered by Mathieu [13,15]. If a 2-dimensional section of the real Leech lattice is considered fixed, then the group $2^{12}M_{24}$ will be reduced to $2^{10}M_{22}$, where $2^{10}$ is the subgroup of $2^{12}$ that doesn't touch the distinguished hexagonal section, while $M_{22}$ is the subgroup of $M_{24}$ that fixes two letters. The subgroup of $0\bullet$ which preserves the structure of the complex Leech lattice is $6.Suz$, where Suz is the Suzuki sporadic group [15], while the subgroup of $6.Suz$ which fixes our choice for the distinguished hexagonal section is $2\times 3^6:M_{11}$, where $M_{11}$ is the smallest Mathieu permutation group.

The miracle of creation in the context of our construction of space-times is that while the group $M_{11}$ acts transitively on the coordinates of complex Leech lattice vectors when 1 coordinate is fixed, the full symmetry group $2 \times 3^6:M_{11}$ has orbits consisting of vectors with shapes that invite association with continuous symmetries similar to those that have already been encountered in supersymmetric unifications of gravity theories of elementary particles in 10-dimensions. For example, there is an orbit consisting of 11-dimensional complex vectors with shape $((2\alpha)^2, \alpha^9)$, where $\alpha$ is a cube root of 1. $2 \times 3^6:M_{11}$ acts on these vectors in the same way that the $11 \times 10/2 = 55$ generators of $SO(11)$ would act on a 2-form gauge field $B_{\mu\nu}$ in 11 dimensions. Therefore it appears that the discrete action of $M_{11}$ within this orbit can be lifted to a continuous $SO(11)$ gauge transformation by exponentiation.

Another interesting orbit of $2 \times 3^6:M_{11}$ emerges from the amusing fact that the Mathieu group $M_{12}$ can be represented as the product of 5-cycle clockwise twists of the vertices of a 3-dimensional icosahedron around each vertex with an antipodal inversion of the icosahedron [13]. This provides a simple way for understanding how $M_{11}$ transforms the complex Leech polytrope with one coordinate fixed. For example, it is immediately evident that the action of $M_{11}$ on the Leech polytrope vectors representing the 2-form gauge field $B_{\mu\nu}$ corresponds to rigid rotations of the icosahedron that fix one axis. It is also straightforward to show that the subgroup of $M_{11}$ corresponding to transformations of the 36 pairs of icosahedron vertices whose 5-cycles leave vertex of the icosahedron fixed can identified with the generators of an SU(3) x $E_6$ gauge symmetry. The fact that the automorphism group of the complex Leech lattice acting within the orbits of complex Leech polytrope with one vertex fixed of can identified with the generators of well known Lie groups allow us to introduce matrix gauge fields to represent the Leech lattice symmetries of our theory. These matrix gauge fields are a natural generalization of the U(1) Chern-Simons gauge field that we used to construct Einstein space-times, and launch us in the direction of pursuing the same sort of strategy for constructing emergent space-times with matter that we used to construct Einstein spaces.

If we ignore the lattice structure of the distinguished hexagonal section of the Leech lattice, then we expect that holomorphic and anti-holomorphic matrix solitons generalizing Eq, (8) will play a central role in our new theory of space-time. Our model follows the Einstein space construction by insisting that the structure of space-time will be determined by a Chern-Simons constraint condition similar to Eq. (6), except the Chern-Simons gauge field is now a complex matrix belonging to the adjoint representation of a Lie algebra The bare bones for our model is the matrix nonlinear Schrodinger equation in 2+1 -dimensions $H\Psi = 0$, where

$$H = \frac{1}{2m} \int d^2x \left( (D_i\Psi)^*(D_i\Psi) + g([\Psi^\dagger, \Psi])^2 \right). \quad (15)$$

The coordinates $x_i$ where $i = 1,2$ refer to the 2-dimensions of the distinguished hexagonal section.

$$D_\alpha = \partial_\alpha + e[A_\alpha, \quad (16)$$

The presence of the gauge potential indicates that our set up is covariant with respect to continuous gauge transformations. Although at this stage it appears that it might be possible to find a generalization of Einstein spaces for any choice of a continuous gauge symmetry, in the following we will focus on the continuous symmetries that arise from the action of $M_{11}$ on the vectors of the complex Leech polytrope. As in the SU(N) case we will assume that the matrix gauge potential is not independent of the wave function, but satisfies the nonabelian Chern-Simons equations:

$$\partial_1 A_2 - \partial_2 A_1 + [A_1, A_2] = -\frac{e}{\kappa}[\Psi^*, \Psi] \quad (17)$$

$$\partial_\alpha A_0 = -\frac{e}{\kappa}\varepsilon_{\alpha\beta}\left(\Psi^* D_\beta \Psi - (D_\beta \Psi^*)\Psi\right)$$

These equations can be thought of as gravity versions of the London equation for a superconductor, and express the "superfluid" nature of the space-time. The constraints (17) can also be introduced by adding a 2 +1 dimensional nonabelian Chern-Simons topological lagrangian:

$$L_{CS} = \frac{\kappa}{2}\varepsilon^{\alpha\beta\gamma}Tr\left(A_\alpha \partial_\beta A_\gamma + \frac{2}{3}A_\alpha A_\beta A_\gamma\right) \quad (18)$$

where $\alpha,\beta,\gamma = 0,1,2$ and the gauge fields are 11x11 matrices. We will also assume that the wave function satisfies self-duality condition $D_\alpha\Psi = \pm i\varepsilon_{\alpha\beta}D_\beta\Psi$. The fact that the self-duality and Chern-Simons constraints for the wave function and gauge fields have the same form as in the SU(N) case discussed in Section 2 means that we can pursue similar strategies for solving these constraints. In particular following the ansatz introduced by Grossman [19] for the wave function in the SU(N) case we propose representing the wave function for space-times with 3-form tensor or $E_6$ vector vacuum fields as a linear combination of raising and lowering operators for the SO(11) or $E_6$ Lie algebras:

$$\Psi = \sum_a e^{\phi_a} E_a^{\pm} \qquad (19)$$

where the index *a* runs over the adjoint representations of SO(11) or $E_6$ as well as SU(3) ( which become 3 "flavor" indices if the SU(3) gauge symmetry is completely broken). The SU(N) nonabelian Chern-Simons constraint condition (6) now becomes the matrix equation

$$\partial_z \partial_{\bar{z}} \phi_a = -\frac{e}{\kappa} \sum_b K_{ab} e^{\phi_b} \qquad (20)$$

where $K_{ab}$ is the Cartan matrix for an SO(11) or SU(3) x $E_6$ Lie algebra associated with the orbits of $M_{11}$. This matrix equation generalizes the classical "Heavenly" equation satisfied by the Kahler potential for a self-dual Einstein space [1]. Just as holomorphic and anti-holomorphic solutions of the classical "Heavenly" equation were used in [3,4] to construct quantum models for Einstein spaces, we also expect that holomorphic and anti-holomorphic solutions of the matrix Eq. 20 can be used to construct quantum models for space-times with 3-form or SU(3) x $E_6$ vacuum fields. How this works in detail for space-times of interest, e.g. gravitational collapse, will be considered in future papers.

We close with a few comments about the general character of our new quantum model for space-time. The appearance of 3-form or Yang-Mills vacuum field strengths in our model for space-time is intriguing from the point of view that these types of background vacuum fields also occur in the ground states of 10-dimensional superstring theories with compacted extra dimensions. It is perhaps worth noting in this connection that the author anticipated some time ago [20] that the massless states in superstring theories might be related to the structure of the Leech lattice. On the other hand superstring theories by themselves don't offer any natural explanation for why ordinary space-time is 3+1 dimensional; whereas in our model 3+1 dimensions plays a special role because of the connection between motion of a charged particle on a hexagonal lattice and

Chern-Simons dynamics. To the author's knowledge hexagonal lattices play no special role in superstring theory.

It follows from Eqs. (17) that the 3-form and Yang-Mills vacuum fields will be strongly by the dynamics of space-time, as well as *visa versa*. Because our self-duality constraint can only be satisfied for some particular configuration of vacuum 3-form and Yang-Mills field strengths, it follows that the vacuum energy will not be zero for general configurations of the vacuum fields. The author and Nick Manton pointed out some time ago [21] that a "geometric" Higgs potential could appear in the 3-dimensions of ordinary space-time as a result of a topologically non-trivial configuration of vacuum gauge fields in extra dimensions. However, the vacuum energy density associated with such a "geometric" Higgs potential would very large unless the charateristic size of the extra dimensions was macroscopic; which would contradict laboratory observations which rule out large extra dimensions. On the other hand, in our model for 3+1 dimensional space-time a change in the vacuum configuration of 3-form or Yang-Mills gauge field strengths can give rise to a positive value for vacuum energy that is consistent with phenomenology if, as suggested in ref. [11], the vacuum energy changes when one crosses a surface where classical general relativity predicts that there should be an event horizon. In this case explanations would be in hand for both the observed magnitude of the cosmological constant and the appearance of a vacuum energy during gravitational collapse.

In summary we believe that the lagrangian field equations (15-18) constitue a first step towards filling the void that arises from the fact that the way matter is treated in classical general relativity doesn't take into account possible changes in the structure of the space-time vacuum. We believe that it is this defect which prevents classical general relativity from providing a physical description of either gravitational collapse or the origin of the big bang.

As a final note we would like to mention that it was Richard Slansky [22] who originally pointed out that in many ways $E_6$ is the nicest Yang-Mills gauge symmetry for a grand unified theory of elementary particles.

Acknowledgements

The author is very grateful for discussions with Bernard Grossman, Robert Laughlin, Pawel Mazur, Emil Mottola, Samuel Braunstein, and Jim Barbieri. This work was performed under the auspices of the U.S. Department of Energy by Lawrence Livermore National Laboratory under Contract DE-AC52-07NA27344.